\documentclass[pre,twocolumn,amsmath,amssymb,floatfix]{revtex4}

\usepackage{graphicx}
\usepackage{dcolumn} 
\usepackage{bm}      
\usepackage{color}

\begin{document}

\title{Scale interactions and scaling laws in rotating flows at moderate
       Rossby numbers and large Reynolds numbers}
\author{P.D. Mininni$^{1,2}$, A. Alexakis$^3$, and A. Pouquet$^2$}
\affiliation{$^1$ Departamento de F\'\i sica, Facultad de Ciencias Exactas y
         Naturales, Universidad de Buenos Aires, Ciudad Universitaria, 1428
         Buenos Aires, Argentina. \\
             $^2$ NCAR, P.O. Box 3000, Boulder, Colorado 80307-3000, U.S.A.\\
             $^3$ Laboratoire Cassiop\'ee, Observatoire de la C\^ote d'Azur,
         BP 4229, Nice Cedex 04, France.}
\date{\today}

\begin{abstract}
The effect of rotation is considered to become important when the 
Rossby number is sufficiently small, as is the case in many geophysical 
and astrophysical flows. Here we present direct numerical simulations 
to study the effect of rotation in flows with moderate Rossby numbers 
(down to $Ro \approx 0.1$) but at Reynolds numbers large enough to observe 
the beginning of a turbulent scaling at scales smaller than the energy 
injection scale. We use coherent forcing at intermediate scales, leaving 
enough room in the spectral space for an inverse cascade of energy to 
also develop. We analyze the spectral behavior of the simulations, the 
shell-to-shell energy transfer, scaling laws, and intermittency, as 
well as the geometry of the structures in the flow. At late times, the 
direct transfer of energy at small scales is mediated by interactions 
with the largest scale in the system, the energy containing eddies 
with $k_\perp \approx 1$, where $\perp$ refers to wavevectors 
perpendicular the axis of rotation. The transfer between modes with 
wavevector parallel to the rotation is strongly quenched. The inverse 
cascade of energy at scales larger than the energy injection scale is 
non-local, and energy is transferred directly from small scales to the 
largest available scale. Also, as time evolves and the energy piles up 
at the large scales, the intermittency of the direct cascade of energy 
is preserved while corrections due to intermittency are found to be the 
same (within error bars) as in homogeneous turbulence.
\end{abstract}
\maketitle

\section{Introduction}
Strong rotation is present in many geophysical and astrophysical flows. 
Its effect is considered to become important when the Rossby number (the 
ratio of the convective to the Coriolis acceleration, or the ratio of the 
rotation period to the eddy turn over time) is sufficiently small. The 
large scales of atmospheric and oceanic flows for example are affected 
by the rotation of the Earth. The Rossby number for mid-latitude synoptic 
scales in the atmosphere is $Ro \approx 0.1$ \cite{Pedlosky}. In the Sun, 
the typical Rossby number in the convective zone is $Ro \approx 0.1-1$ 
\cite{Miesch00}. Furthermore, the Reynolds number ($Re$, the ratio of the 
convective to the viscous acceleration) in these systems is also very 
large, and the flows are in a turbulent state.

Besides rotation, stratification is also important in the atmosphere, 
the ocean, and other geophysical flows. Many studies have considered 
solely the effect of rotation in a turbulent flow, as a first step to 
gain better understanding of the fluid dynamics of geophysical systems. 
For rapid rotation (very small Rossby numbers), significant progress 
has been made by applying resonant wave theory 
\cite{Greenspan,Waleffe93,Embid96,Babin96} and weak turbulence theory 
\cite{Galtier03}. In these approaches, the flow is considered as a 
superposition of inertial waves with a short period, and the evolution 
of the system for long times is derived considering the effect of 
resonant triad interactions. This explains successfully the observed 
enhanced transfer of energy from the small to the large scales 
\cite{Smith05}, and sheds light on the mechanism that drives the flow to be
quasi-two dimensional at large scales \cite{Waleffe93,Smith99}. Energy 
in three dimensional modes is transferred by a subset of the resonant 
interactions to modes with smaller vertical wavenumber. Spectral 
closures \cite{Cambon89,Cambon97} give similar results.

However, resonant wave theories are only valid when the rotation period 
is much shorter than the eddy turnover time at all scales. For large 
Reynolds numbers, small scales are excited with a characteristic 
timescale proportional to the eddy turnover time, that decreases as 
the scales become smaller. Therefore the approximations made in such 
theories can break down at sufficiently small scales, provided that the 
Reynolds number is large enough for these scales to be excited. How the 
results of resonant wave theory extend to the case of only moderate 
Rossby numbers but very large Reynolds numbers is still unclear. Several 
phenomenological theories have been developed to consider the case with 
large $Re$ (see e.g. \cite{Zeman94,Zhou95,Canuto97,Muller07}) leading to
different results for the scaling of the energy spectrum.

In numerical simulations, the study of rotating turbulent flows is
constrained by the computational cost of properly resolving the
inertial waves and the resonant triadic interactions, together with the cost
of resolving the small scale fluctuations when $Re$ is large. Inverse
cascades were shown to develop and anisotropies to appear in low
resolution ($32^3$ and $64^3$ grid points) simulations
\cite{Bardina85,Hossain94,Bartello94}, either solving the
equations of motion directly or using a subgrid model. Small aspect
ratio boxes were considered in \cite{Smith96,Smith99} allowing for an increase
in resolution. Simulations at higher resolution were done later by
\cite{Yeung98} studying in particular the behavior of the shell-to-shell 
energy transfer. Recently, simulations with large Reynolds number and
small Rossby number were performed using $128^3$ grids and 8th-order
hyperviscosity \cite{Chen05}, thus confirming the dominant role of
resonant triads for rapid rotation at large $Re$, although the results
also suggest that resonant wave theories can be valid only for a finite
interval of time. All these simulations also give different results
for the scaling of the energy spectrum at scales larger than the
forcing scale; it was shown in \cite{Smith05}, using a truncated model, that
this can be the result of how all the relevant timescales are resolved.

Here we study the effect of rotation in a turbulent flow in high
resolution direct numerical simulations (up to $512^3$ grid points).
Simulations at this resolution were also performed recently in
\cite{Muller07}; in this case, energy was injected at the
largest scale available and the focus was solely on the
scaling of small-scale fluctuations, showing depletion of the energy 
cascade and reduced intermittency. Our main objective, on the other 
hand, is to study the
statistical properties of the fluctuations in flows with moderate
Rossby numbers (down to $Ro \approx 0.1$) but at Reynolds numbers
large enough to observe the beginning of a turbulent scaling at
scales smaller than the energy injection scale. To this end, we use
coherent forcing at intermediate scales, leaving enough room in the
spectral space for an inverse cascade of energy to develop when the
Rossby number is small enough. We also use the largest value of the
Reynolds number allowed by our grid to observe a direct transfer of energy
at small scales. After describing the simulations, we study its
spectral behavior, the shell-to-shell energy transfer, scaling laws
and intermittency, and finally the geometry of the structures in the
flow.

\section{Numerical simulations}

We solve numerically the equations for an incompressible rotating fluid
with constant mass density,
\begin{equation}
\frac{\partial {\bf u}}{\partial t} + \mbox{\boldmath $\omega$} \times
    {\bf u} + 2 \mbox{\boldmath $\Omega$} \times {\bf u}  =
    - \nabla {\cal P} + \nu \nabla^2 {\bf u} + {\bf F} ,
\label{eq:momentum}
\end{equation}
and
\begin{equation}
\nabla \cdot {\bf u} =0 ,
\label{eq:incompressible}
\end{equation}
where ${\bf u}$ is the velocity field,
$\mbox{\boldmath $\omega$} = \nabla \times {\bf u}$ is the vorticity,
${\cal P}$ is the total pressure (modified by the centrifugal term)
divided by the mass density, and $\nu$ is the kinematic viscosity.
Here, ${\bf F}$ is an external force that drives the turbulence, and
we chose the rotation axis to be in the $z$ direction:
$\mbox{\boldmath $\Omega$} = \Omega \hat{z}$, with $\Omega$ the
rotation frequency.

The mechanical forcing ${\bf F}$ is given by the Taylor-Green (TG)
flow \cite{Taylor37}
{\setlength\arraycolsep{2pt}
\begin{eqnarray}
{\bf F} &=& F_0 \left[ \sin(k_0 x) \cos(k_0 y)
     \cos(k_0 z) \hat{x} - \right. {} \nonumber \\
&& {} \left. - \cos(k_0 x) \sin(k_0 y)
     \cos(k_0 z) \hat{y} \right] ,
\label{eq:TG}
\end{eqnarray}}
\noindent where $F_0$ is the forcing amplitude. Although the forcing injects
energy directly only into the $x$ and $y$ components of the flow, in
the absence of rotation ($\Omega = 0$) the resulting flow is fully
three-dimensional because of pressure gradients that excite the $z$
component of the velocity \cite{Taylor37,Morf80}. The resulting flow
has no net helicity, although locally regions with strong positive
and negative helicity develop. It is also worth noting that this 
forcing injects zero energy in the $k_z=0$ modes, whose amplification 
observed in the strongly rotating cases is only due to a cascade 
process.

\begin{table}
\caption{\label{table:runs}Parameters used in the simulations. $N$ is
         the linear grid resolution, $k_0$ the wavenumber used in the
         forcing, $\nu$ the kinematic viscosity, $\Omega$ the rotation
         rate, $t_\textrm{max}$ the maximum number of turnover times
         computed; $Re, \ Ro, \ $ and $Ek$ are respectively the 
         Reynolds, Rossby and Ekman numbers.}
\begin{ruledtabular}
\begin{tabular}{ccccccccc}
Run& $N$&$k_0$&     $\nu$       &$\Omega$&$t_\textrm{max}$ & $Re$ & $Ro$ &
  $Ek$  \\
\hline
A1 & 256&  2  &$2\times 10^{-3}$& $0.08$ &  $45$  & $900$  & $4.50$ & 
  $5\times10^{-3}$ \\
A2 & 256&  2  &$2\times 10^{-3}$& $0.40$ &  $45$  & $900$  & $0.70$ & 
  $8\times10^{-4}$ \\
A3 & 256&  2  &$2\times 10^{-3}$& $0.80$ &  $45$  & $900$  & $0.35$ & 
  $4\times10^{-4}$ \\
A4 & 256&  2  &$2\times 10^{-3}$& $1.60$ &  $45$  & $900$  & $0.17$ & 
  $2\times10^{-4}$ \\
A5 & 256&  2  &$2\times 10^{-3}$& $3.20$ &  $150$ & $900$  & $0.09$ & 
  $1\times10^{-4}$ \\
A6 & 256&  2  &$2\times 10^{-3}$& $8.00$ &  $185$ & $900$  & $0.03$ & 
  $3\times10^{-5}$ \\
B1 & 512&  4  &$8\times 10^{-4}$& $0.40$ &  $17$  & $1100$ & $1.40$ & 
  $1\times10^{-3}$ \\
B2 & 512&  4  &$8\times 10^{-4}$& $1.60$ &  $25$  & $1100$ & $0.35$ & 
  $3\times10^{-4}$ \\
B3 & 512&  4  &$8\times 10^{-4}$& $8.00$ &  $40$  & $1100$ & $0.07$ & 
  $6\times10^{-5}$ \\
\end{tabular}
\end{ruledtabular}
\end{table}

Two sets of runs were done at resolutions of $256^3$ (set A) and
$512^3$ grid points (set B). The parameters for all the runs are listed
in Table \ref{table:runs}. With Taylor-Green forcing, the spherical
shell in Fourier space where energy is injected has wavenumber
$k_F = \sqrt{3} k_0$, or equivalently, at a scale
$L_F = 2 \pi/k_F$. For the runs in set A, $k_F \approx 3.5$, and
for the runs in set B, $k_F \approx 6.9$; as a result, there is more
room in spectral space for an inverse cascade to take place in the B 
runs.

All the runs in set A were started from a fluid at rest. At $t=0$, the
rotation and the external forcing were switched on, until reaching a
turbulent steady state, or until an inverse cascade was well developed
in the case of large rotation rates. The runs in set B were done as
follows. Run B1 was started from a fluid at rest and after turning on
the rotation and external forcing, the run was continued to reach a
turbulent steady state. Runs B2 and B3 were started from a snapshot
of the velocity field from the steady state of run B1, and both
runs were continued until a new steady state was reached, or an
inverse cascade developed. This latter method proved useful in saving
computing time, as no differences were observed when comparing the
late time evolution of the runs in the two sets. In all simulations, 
a dissipative range was properly resolved, and the time step was much 
smaller than all the relevant timescales.

\begin{figure}
\includegraphics[width=8cm]{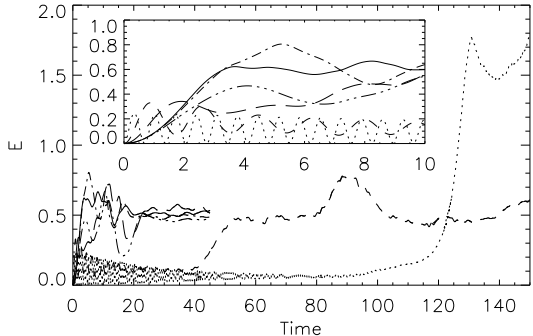}
\caption{Time history of the energy for set A: A1 (solid),
    A2 (dash-dot), A3 (dash-triple dot), A4 (long dash),
    A5 (dash), and A6 (dot); the inset shows a detail of the
    evolution at early times when waves prevail. Note the large 
    increase in energy as $Ro$ decreases.}
\label{fig:timee} \end{figure}

We define the integral and Taylor scales of the flow respectively as
\begin{equation}
L = 2\pi \frac{\int{E(k) k^{-1} dk}}{\int{E(k) dk}},
\label{eq:integral}
\end{equation}
and
\begin{equation}
\lambda = 2\pi \left(\frac{\int{E(k) dk}}{\int{E(k) k^2 dk}}\right)^{1/2},
\label{eq:taylor}
\end{equation}
where $E(k)$ is the energy spectrum. Since for large $\Omega$ an inverse
cascade develops, these two scales are useful to describe the evolution 
of characteristic scales in the flow with time. However, to avoid a time 
dependence of the Reynolds and Rossby numbers, we define for each run 
the Reynolds number as
\begin{equation}
Re = \frac{L_F U}{\nu} ,
\end{equation}
and the Rossby number as
\begin{equation}
Ro = \frac{U}{2 \Omega L_F} .
\end{equation}
We also define the Ekman number as
\begin{equation}
Ek = \frac{Ro}{Re} = \frac{\nu}{2 \Omega L_F^2} .
\end{equation}
The turnover time at the forcing scale is then defined as $T = L_F/U$
where $U=\left< u^2 \right>$ is the r.m.s. velocity measured in the
turbulent steady state, or when the inverse cascade starts. The amplitude 
of the forcing
 $F_0$ in the simulations is increased as $\Omega$ is increased in order to
have $U \approx 1$ in all the runs.

\begin{figure}
\includegraphics[width=8cm]{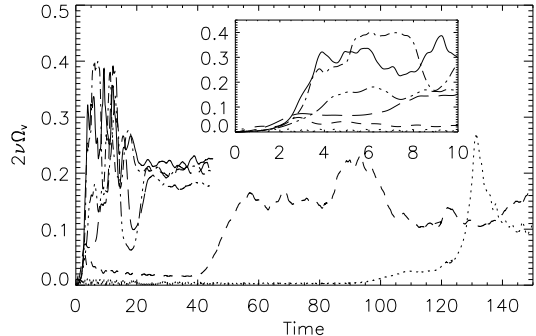}
\caption{Time history of the energy dissipation rate (labels as in
    Fig. \ref{fig:timee}) the inset shows the evolution at early times.}
\label{fig:timed} \end{figure}

\section{Time evolution and spectra}

Figure \ref{fig:timee} shows the time history of the energy in the
runs in set A. Runs A1-A4 show a similar evolution, but runs A5 and
A6 evolve differently. As the Rossby number decreases, a transient
develops in which the total energy oscillates with a frequency that
increases with $\Omega$. It takes a longer time for this transient to
decay as $\Omega$ increases, and then the energy increases suddenly 
and a turbulent regime develops. An inverse cascade of energy is 
observed in run A6 after $t \approx 120$. The increase in the energy 
observed after this time is also accompanied by a monotonous increase 
with time of the flow integral scale $L$. Even in the runs in set B, 
that are restarted from a pre-existing turbulent steady state, long 
runs are needed to reach another turbulent state after turning on the 
rotation. As an example, in run B3 it takes $\approx 20$ turnover 
times for the transient to decay and for an inverse cascade of energy 
to develop.

The energy dissipation rate $2\nu  \int \omega^2/2 dV$ as a function of 
time is shown in Fig. \ref{fig:timed}.
As the Rossby number decreases, the peak of the dissipation rate is
reached at later times, and then it saturates. Note that during the 
early transient in runs A5 and A6, the dissipation is almost negligible, 
while in the saturated state the mean dissipation rate decreases slowly 
with decreasing Rossby number.

\begin{figure}
\includegraphics[width=8cm]{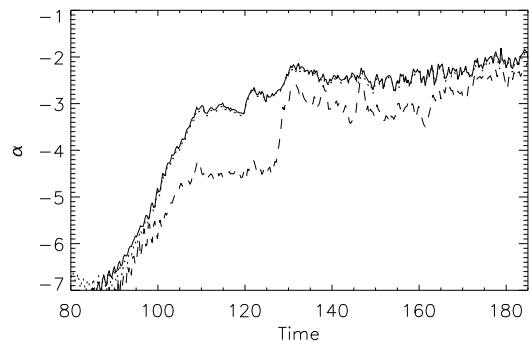}
\caption{Spectral index $\alpha$ as a function of time in run A6, in the 
    isotropic energy spectrum $E(k)$ (solid), in the $E(k_\perp)$ spectrum 
    (dot), and in the $E(k_\parallel)$ spectrum (dash). Note that the 
    energy is dominated by the orthogonal modes.}
\label{fig:index} \end{figure}

The shape of the energy spectrum evolves with time, specially after the 
transient as turbulence sets in, and later again as the spectrum becomes 
dominated by the contribution from the largest scales when the Rossby 
number is small enough for an inverse cascade to develop. Figure 
\ref{fig:index} shows the time evolution of the spectral index $\alpha$ 
(the exponent in the region of the spectrum with $k>k_F$ that follows a 
power law $\sim k^\alpha$) in run A6. Three curves are shown, which 
correspond respectively to the spectral index computed in the isotropic 
energy spectrum $E(k)$, in the perpendicular energy spectrum 
$E(k_\perp)$ (where 
$k_\perp$ denotes the wavevectors perpendicular to 
$\mbox{\boldmath $\Omega$}$), and the parallel spectrum $E(k_\parallel)$ 
(where $k_\parallel$ denotes the wavevectors parallel to 
$\mbox{\boldmath $\Omega$}$).

\begin{figure}
\includegraphics[width=8cm]{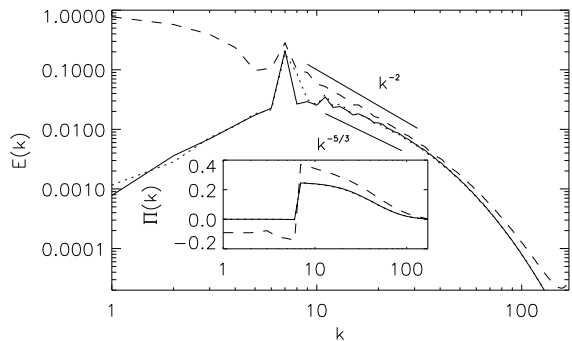}
\caption{Isotropic energy spectra at late times in runs B1 (solid, 
    $t \approx 16$), B2 (dot, $t \approx 24$), and B3 (dash, 
    $t \approx 40$) at low $Ro$. Two slopes are given as a reference. The 
    inset shows the isotropic energy flux for the same runs.}
\label{fig:spectrum} \end{figure}

Before $t \approx 80$, we cannot recognize a power law in the energy 
spectra. After $t \approx 80$, the spectral indices in $E(k)$ and $E(k_\perp)$ 
grow monotonically from a value of $-7$ until reaching a plateau with 
$\alpha \approx -3$ at $t \approx 110$. The energy spectra $E(k)$, 
$E(k_\perp)$, and $E(k_\parallel)$ show wide and steep power law behavior from 
$t \approx 80$ to $t \approx 120$. During this transient, the energy 
flux is almost zero, as can also be expected from the small value of the 
energy dissipation in run A6 before $t\approx 110$ (Fig. \ref{fig:timed}). 
The end of the transient at $t \approx 110$ and the plateau in $\alpha$
correspond respectively to the increase in the energy and in the 
energy dissipation 
rate showed in Figs. \ref{fig:timee} and \ref{fig:timed}. The spectral 
index in $E(k_\parallel)$ also has a plateau with $\alpha \approx -4.5$. 
However, as the inverse cascade sets in and the energy piles up at the 
largest available scale in the system, the spectral index changes again 
and seem to slowly evolve towards $\alpha \approx -2$ in both $E(k)$ and 
$E(k_\perp)$.

\begin{figure}
\includegraphics[width=8cm]{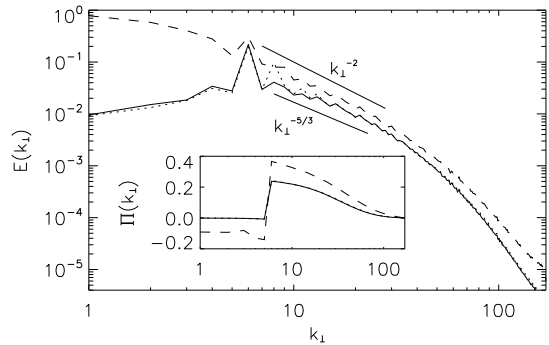}
\caption{$E(k_\perp)$ and $\Pi(k_\perp)$ (inset) at late times in runs
    B1, B2, and B3. Labels are as in Fig. \ref{fig:spectrum}.}
\label{fig:spectper} \end{figure}

Note that the inverse cascade only starts after $\approx 10$ turnover 
times after the turbulent state is reached at $t \approx 110$. This 
can be understood as follows. The energy spectrum observed before 
$t \approx 110$ has almost no flux. Nonlinear transfer of energy is 
required for the flow to become two-dimensional under the effect of 
rotation \cite{Waleffe93,Cambon97,Smith99}, and the nonlinear transfer 
is negligible until $t \approx 110$. Then, after a few turnover times, 
the flow undergoes a transition and the inverse cascade sets in.

The early transient is only observed in the runs in set A, since the runs 
in set B are started from a turbulent steady state. However, after the 
transient the spectral evolution of the runs in set A and B is similar. 
Since runs in set B have more scale separation for an inverse cascade to 
develop when $Ro$ is small enough, we focus now on this set of runs. 
We show in Fig. \ref{fig:spectrum} the isotropic energy spectrum at late 
times in runs B1-B3. While runs B1 and B2 show no growth of energy at scales
larger than the mechanical forcing, except for some backscattering with
a $\sim k^2$ spectrum, run B3 at late times is dominated by the energy
in the $k=1$ shell. At scales smaller than the forcing scale, the
spectrum of run B3 is steeper than that of runs B1 and B2, and compatible
with a $\sim k^{-2}$ scaling. The inset in Fig. \ref{fig:spectrum} shows
the isotropic energy flux in the same runs. Note that in run B3, the
flux at scales larger than the forcing scale is negative and
approximately constant, indicating the development of an inverse cascade
of energy for small $Ro$. At smaller scales, the energy flux is positive. 
We thus conclude that in rotating flows, both the direct and inverse 
energy cascades can cohabit.

\begin{figure}
\includegraphics[width=8cm]{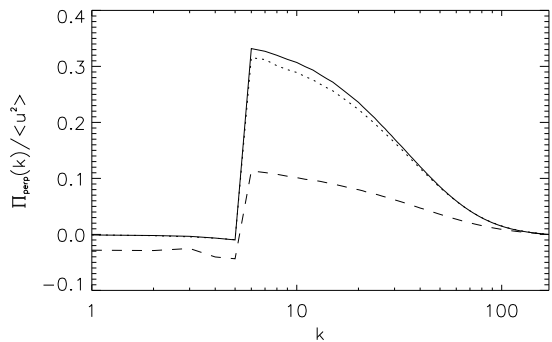}
\caption{$\Pi(k_\perp)/\left< u^2 \right>$ at late times in runs B1, B2, 
    and B3. Labels are as in Fig. \ref{fig:spectrum}.}
\label{fig:fluxper} \end{figure}

The energy spectrum $E(k_\perp)$ is shown in Fig. \ref{fig:spectper}, 
together with the energy flux $\Pi(k_\perp)$. The spectrum and flux are 
similar to the isotropic ones (indicating most of the energy is in these 
modes), and $\Pi(k_\perp)$ confirms the development of an inverse 
cascade of energy in $k_\perp$ at scales larger than the forcing scale 
in run B3, and a direct cascade at smaller scales with a 
$\sim k_\perp^{-2}$ scaling. Figure \ref{fig:fluxper} shows the 
energy flux $\Pi(k_\perp)$ normalized by the r.m.s. velocity in each 
run. Note that the increase of the flux observed in the inset of 
Fig. \ref{fig:spectper} is only due to the increase in the energy of 
the system as the inverse cascade piles up energy at the largest 
available scale. As Fig. \ref{fig:fluxper} indicates, the actual transfer 
of energy is slowed down by the rotation, and run B3 shows a smaller 
normalized flux than the other two runs at scales smaller than the 
forcing scale.

\begin{figure}
\includegraphics[width=8cm]{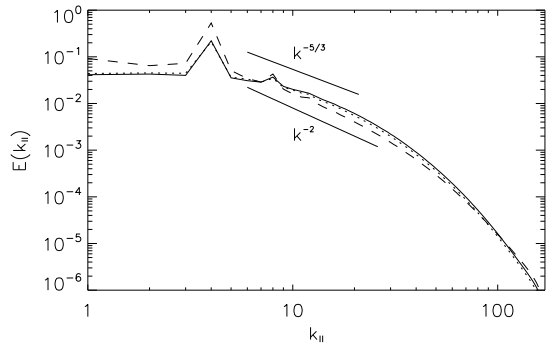}
\caption{$E(k_\parallel)$ at late times in runs B1, B2, and B3. Labels
    are as in Fig. \ref{fig:spectrum}.}
\label{fig:spectpar} \end{figure}

On the other hand, there is no clear scaling in the small scales in 
$E(k_\parallel)$, nor an inverse cascade at large scales (see Figure 
\ref{fig:spectpar}). The $E(k_\parallel)$ spectrum in run B3 is 
steeper than the $E(k_\perp)$ spectrum, consistent with the results 
shown in Fig. \ref{fig:index} for run A6 at late times. Slopes 
$\sim k^{-5/3}$ and $\sim k^{-2}$ are shown in Fig. \ref{fig:spectpar} 
only as a reference.

\section{Energy transfer}

In this section we study the scale interactions and energy transfer in
rotating turbulent flows. A study of the energy transfer in this context, 
albeit at lower resolution, was done before by \cite{Yeung98}. We will 
focus on runs B1, B2, and B3, that have enough scale separation for 
direct and inverse cascades to develop when $Ro$ is small enough. Similar 
results were obtained in the analysis of the runs in set A.

\begin{figure}
\includegraphics[width=8cm]{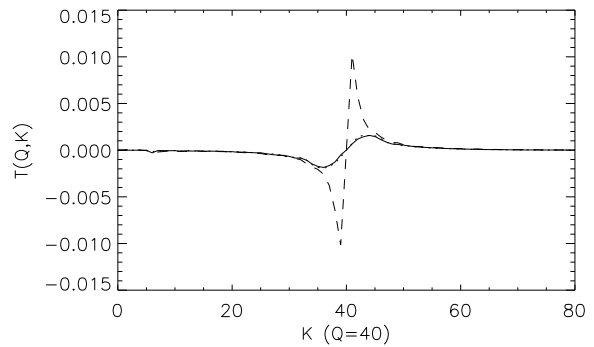}
\caption{Shell-to-shell transfer function $T(Q,K)$ at $Q=40$ for runs
    B1 (solid), B2 (dot), and B3 (dash) at late times.}
\label{fig:transfer} \end{figure}

To investigate the transfer of energy among different scales we consider
the shell filter decomposition of the velocity field,
\begin{equation}
{\bf u}({\bf x}) = \sum_K \tilde{\bf u}_K ({\bf x}) ,
\end{equation}
where $K$ denotes a foliation of Fourier space in shells, that for our
purposes can be taken as spheres 
\cite{Alexakis05,Mininni05,Alexakis05b,Mininni06}
\begin{equation}
{\bf u}_K ({\bf x}) = \sum_{K \le |{\bf k}| \le K+1} \tilde{\bf u}_{\bf k}
    e^{i{\bf k}\cdot \bf{x}} ,
\end{equation}
cylinders \cite{Alexakis07}
\begin{equation}
{\bf u}_{K_\perp} ({\bf x}) = \sum_{K \le |{\bf k_\perp}| \le K+1}
    \tilde{\bf u}_{\bf k} e^{i{\bf k}\cdot \bf{x}} ,
\end{equation}
or planes \cite{Alexakis07}
\begin{equation}
{\bf u}_{K_\parallel} ({\bf x}) = \sum_{K \le |{\bf k_\parallel}| \le K+1}
    \tilde{\bf u}_{\bf k} e^{i{\bf k}\cdot \bf{x}} .
\end{equation}
Then, we can define the shell-to-shell transfer between these shells as
\begin{equation}
T(Q,K) = -\int {\bf {u}_K (u \cdot \nabla) {u}_Q  } \, d{\bf x}^3 .
\end{equation}
This function expresses the transfer rate of energy lying in the shell
$Q$ to energy lying in the shell $K$. It satisfies the symmetry property
$T(Q,K) = -T(K,Q)$ \cite{Alexakis05}, and the numbers labeling the shells
$Q$ and $K$ can correspond to any of the foliations of Fourier space
listed above \cite{Alexakis07}. In particular, we will study the cases
$T(Q,K)$, $T(Q_\perp,K_\perp)$, and $T(Q_\parallel,K_\parallel)$. The
energy fluxes discussed in the previous section can be reobtained in terms
of the shell-to-shell transfer function as
\begin{equation}
\Pi(k) = - \sum_{K=0}^k \sum_Q T(Q,K) ,
\end{equation}
where again the wavenumbers $k$, $K$, and $Q$ can correspond to 
different foliations of Fourier space depending on the subindex.

Note that for the definition of the shells a linear binning is used. 
Alternatively, the shells can be defined by a logarithmic binning of 
spectral space with intervals ($\gamma^n K_0, \gamma^{n+1}K_0$] for 
some positive $\gamma> 1$ and for integer $n$. However, logarithmic 
binning cannot distinguish transfer between linearly spaced neighbor 
shells (from the shell $K$ to the shell $K+1$) from the transfer 
between logarithmic neighbor shells (from $K$ to $\gamma K$). If 
the cascade is the result of interactions with the large-scale flow 
(e.g., with modes with wavenumber $k_F$ associated to the external 
forcing), the energy in a shell $K$ will be transferred to the shell 
$K + k_F$. Logarithmic binning does not distinguish this transfer from 
the transfer due to local triadic interactions that transfer the energy 
from $K$ to $\gamma K$. For this reason we use linear binning, but we 
note that care needs to be taken when using the word ``scale'' that 
implies in general a logarithmic division of the spectral space. The 
transfer among logarithmic shells can be reconstructed at any time later 
by summing over the linearly spaced shells.

\begin{figure}
\includegraphics[width=8cm]{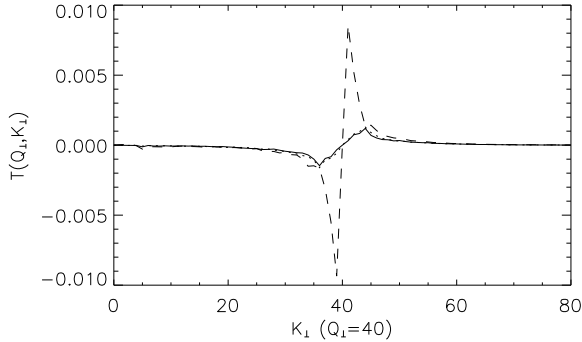}
\caption{Shell-to-shell transfer function $T(Q_\perp,K_\perp)$ at 
    $Q_\perp=40$ for runs B1, B2, and B3. Labels are as in Fig. 
    \ref{fig:transfer}.}
\label{fig:transper}
\end{figure}

\begin{figure}
\includegraphics[width=8cm]{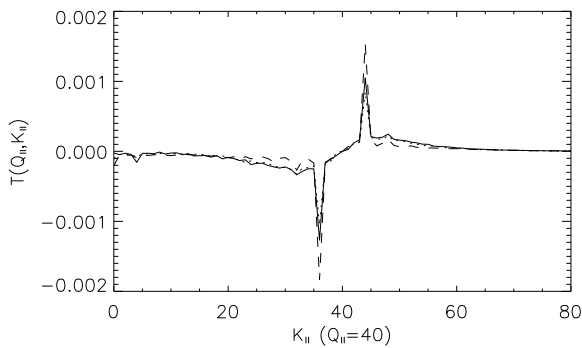}
\caption{Shell-to-shell transfer function $T(Q_\parallel,K_\parallel)$
    at $Q_\parallel=40$ for runs B1, B2, and B3. Labels are as in Fig.
    \ref{fig:transfer}. Notice these transfers are roughly 5 times 
    weaker than in the $\perp$ case.}
\label{fig:transpar}
\end{figure}

Figure \ref{fig:transfer} shows the shell-to-shell transfer $T(Q,K)$ at
$K=40$ for runs B1, B2, and B3 at late times. The negative peak to the 
left indicates energy is transfered from these $K$-shells to the shell 
$Q=40$, while the positive peak to the right indicates energy goes from 
the $Q=40$ shell to those $K$-shells. In runs B1 and B2 the shell-to-shell 
transfer peaks at $|Q-K| \approx k_F$. This was observed before in 
simulations of homogeneous turbulence \cite{Alexakis05b,Mininni06}, and 
indicates that the energy transfer is local (the energy goes from a shell 
$Q$ to a nearby shell $K$, although the step in the energy cascade is 
independent of that scale and related to the forcing scale). However, 
in run B3 the transfer strongly peaks at $|Q-K| \approx 1$. The same 
effect is observed in $T(Q_\perp,K_\perp)$ shown in Fig. \ref{fig:transper}. 
This indicates that at late times in run B3, the direct transfer of 
energy at small scales is mediated by interactions with the largest scale 
in the system, the energy containing eddies with $k_\perp \approx 1$ (see 
Fig. \ref{fig:spectper}). As a result, the timescale associated with the 
direct cascade of energy in $k_\perp$ increases (and its flux reduces, 
see Fig. \ref{fig:fluxper}), since the energy is transfered in smaller 
steps in Fourier space than in the case of the B1 and B2 runs.

\begin{figure*}
\includegraphics[width=17cm]{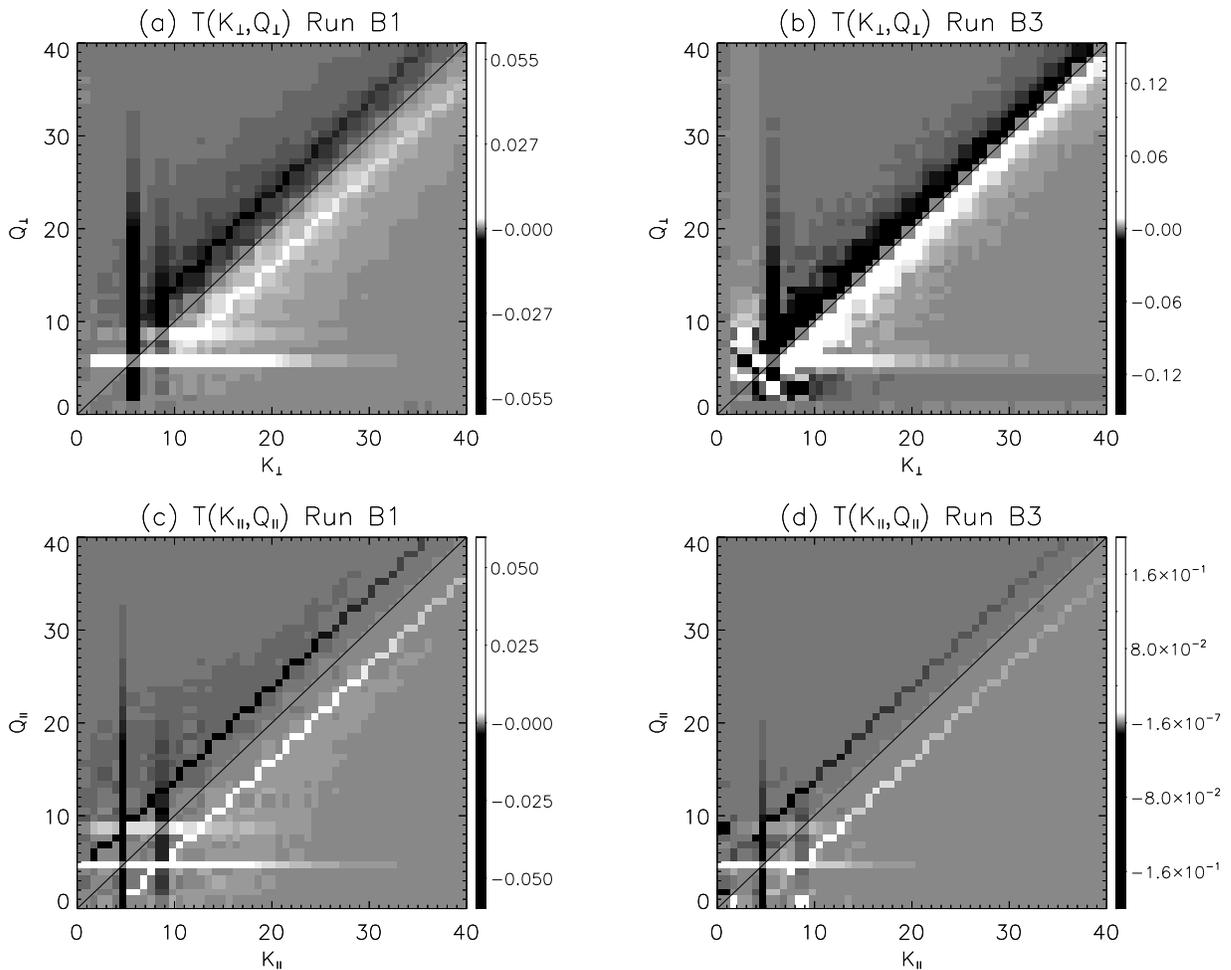}
\caption{Shell-to-shell energy transfer functions $T(Q_\perp,K_\perp)$ (a,b)
    and $T(Q_\parallel,K_\parallel)$ (c,d) at late times in runs B1 (a,c) 
    and B3 (b,d). Notice the quenching of the transfer in case (d), except 
    for the interactions with the forcing scale.}
\label{fig:alltrans}
\end{figure*}

The shell-to-shell transfer $T(Q_\parallel,K_\parallel)$ at
$Q_\parallel=40$ for the same runs is shown in Fig. \ref{fig:transpar}.
The dependence with the Rossby number of this transfer function is less
drastic. In all runs, the transfer function $T(Q_\parallel,K_\parallel)$ 
peaks at $|Q_\parallel-K_\parallel| \approx k_F$. Since there is no 
inverse cascade of energy in $k_\parallel$, the energy containing scale 
in this direction does not change as the Rossby number is decreased, and 
neither does the position of the peaks in $T(Q_\parallel,K_\parallel)$. 
However, note the drop in the amplitude of the transfer in run B3 for 
all shells except the ones satisfying $|Q_\parallel-K_\parallel| = k_F$. 
As a result, for small Rossby number the transfer of energy between shells 
with $Q_\parallel$ and $K_\parallel$ is quenched except for the direct 
interactions with the external forcing. Most of the interactions 
responsible for the transfer of energy to small scales between different 
$k_\parallel$ shells are then interactions with the forcing.

Figure \ref{fig:alltrans} shows the transfer functions $T(Q_\perp,K_\perp)$ 
and $T(Q_\parallel,K_\parallel)$ in runs B1 and B3 for all values of $K$ and
$Q$ up to 40. In all cases, the white and black bands near 
$K \approx k_F$ and $Q \approx k_F$ indicate a small amount of energy 
injected by the external forcing that is directly transfered to all 
wavenumbers up to $\approx 30$. For $K$ and $Q$ larger than $k_F$, the 
figures confirm the results of the direct cascade of energy presented in 
Figs. \ref{fig:transper} and \ref{fig:transpar}. For wavevectors 
perpendicular to $\mbox{\boldmath $\Omega$}$, as the Rossby number is 
decreased the peaks in $T(Q_\perp,K_\perp)$ move closer to the diagonal 
$K_\perp=Q_\perp$ [Figs. \ref{fig:alltrans}(a) and (b)], indicating the 
direct cascade in the perpendicular direction takes place in smaller 
k-steps given by the largest scale of the system. For all wavevectors, the 
energy in the parallel direction [see $T(Q_\parallel,K_\parallel)$ in 
Figs. \ref{fig:alltrans}(c) and (d)] is transfered to smaller scales, 
and the cascade step does not depend on the Rossby number. However, all 
transfer except the transfer with $|Q_\parallel-K_\parallel| = k_F$ is 
strongly quenched in run B3.

\begin{figure}
\includegraphics[width=8cm]{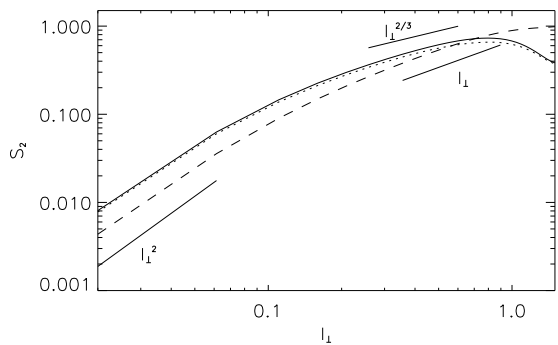}
\caption{Second order longitudinal structure function $S_2(\ell_\perp)$
(where $\ell_\perp$ denotes increments were taken in the direction
perpendicular to $\mbox{\boldmath $\Omega$}$) for runs B1 (solid), B2
(dot), and B3 (dash).}
\label{fig:struct2} \end{figure}

The development of a non-local inverse transfer can be observed in Fig. 
\ref{fig:alltrans}(b) for $K_\perp < k_f$ and $Q_\perp < k_f$. The 
transfer is inverse, since below the diagonal $Q_\perp=K_\perp$ regions 
with negative (dark gray and black) $T(Q_\perp,K_\perp)$ can be observed. 
This means that energy is taken from e.g., $K_\perp = 20$ and transfered 
to shells with $Q_\perp < k_F$. The transfer is also non-local, since 
this inverse transference takes place between disparate scales. The 
non-local transfer of energy in rotating turbulence shares similarities 
with the inverse cascade of magnetic helicity in magnetohydrodynamics 
(MHD) \cite{Alexakis06,Alexakis07b}. Near the diagonal $Q_\perp=K_\perp$ 
the transfer is more complex. The inverse transfer superposes with a 
(smaller in net amplitude) direct local transfer (dark spots below and 
near the diagonal, and light spots above and near it, for $K_\perp$ and 
$Q_\perp$ smaller than $k_F$). This small direct transfer of energy at 
large scales is the result of a reflection of energy at $K=1$, and was also 
observed in studies of the inverse cascade of magnetic helicity in MHD 
\cite{Alexakis06}. The reflection of energy in Fourier space when it 
reaches the largest scale in the box suggests that the late time evolution 
can be dependent on the boundary conditions, a property that was already 
observed in simulations of two dimensional turbulence 
\cite{McWilliams84,Brachet86,Babiano87,Benzi87,Maltrud91}. In our case, 
the simulations do not contain a large-scale dissipation mechanism 
(such as a hypo-viscosity), and therefore energy piles up at the 
largest available scale until its growth is stopped by the (small-scale) 
dissipation.

\section{Scaling laws and intermittency}

\begin{figure}
\includegraphics[width=8cm]{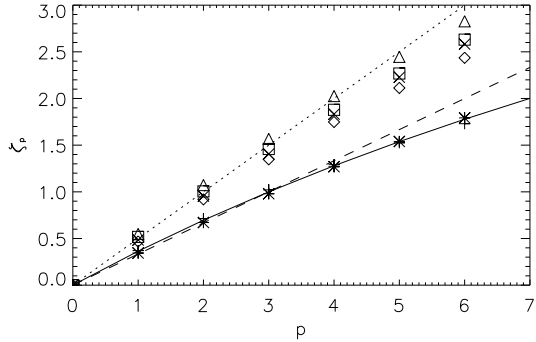}
\caption{Scaling exponents $\zeta_p$ for the steady state of runs B1 ($+$),
    B2 ($*$), and for run B3 at different times: $t \approx 20$
    ($\Diamond$), $t \approx 25$ ($\times$), $t \approx 30$ ($\Box$),
    and $t \approx 40$ ($\triangle$). The solid line corresponds to the
    scaling exponents given by the She-L\'ev\^eque model \cite{She94},
    the dash line is the Kolmogorov prediction $\zeta_p = p/3$, and
    the dotted line is $\zeta_p = p/2$.}
\label{fig:exponent}
\end{figure}

In this section, we consider the anisotropic inertial range scaling of the 
runs in Table \ref{table:runs} as described by the longitudinal velocity
increments in the direction perpendicular to rotation,
\begin{equation}
\delta u({\bf x},\ell_\perp)  = \hat{\bf r} \cdot
    \left[{\bf u}({\bf x}+\ell \hat{\bf r}) - {\bf u}({\bf x}) \right] ,
\end{equation}
where $\hat{\bf r}$ is a unit vector perpendicular to
$\mbox{\boldmath $\Omega$}$. The longitudinal structure functions
$S_p(\ell_\perp)$ (with displacements along $\ell_{\perp}$)
can then be defined as
\begin{equation}
S_p(\ell_\perp) = \left< \delta u({\bf x},\ell_\perp) \right>,
\end{equation}
where the brackets denote spatial average. If the flow is self-similar,
we expect $S_3(\ell_\perp) \sim {\ell_\perp}^{\zeta_p}$, where $\zeta_p$
are the scaling exponents. In isotropic and homogeneous hydrodynamic
turbulence, the K\'arm\'an-Howarth theorem implies $S_3(\ell) \sim \ell$,
and the Kolmogorov energy spectrum follows from the assumption
$S_3(\ell) \sim \ell^{p/3}$ \cite{Frisch}. In practice, the spontaneous
development of strong gradients in the small scales of a turbulent
flow gives rise to corrections to this scaling, a phenomenon referred
to as intermittency.

\begin{figure}
\includegraphics[width=8cm]{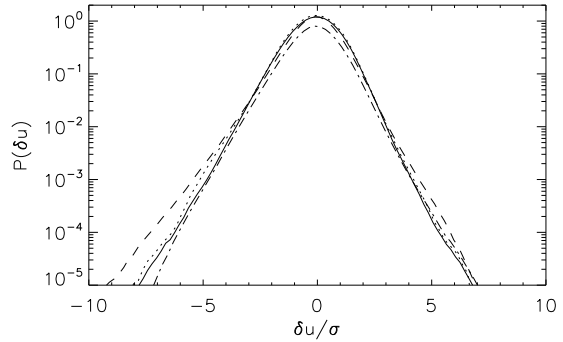}
\caption{Pdf of the longitudinal velocity increments ($\ell_\perp=3\eta$)
    for run B3 at different times: $t \approx 20$ (solid), $t \approx 25$
    (dot), $t \approx 30$ (dash), and $t \approx 40$ (dash-dot); $\eta$ 
    is the Kolmogorov dissipative length.}
\label{fig:histo01}
\end{figure}

From dimensional analysis, if the energy spectrum at small scales in
rotating turbulence is $E \sim k_\perp^{-2}$, we expect 
$S_2 \sim \ell_\perp$. Figure \ref{fig:struct2} shows the second order
structure function for runs B1, B2, and B3 at late times outside the 
wave regime when the turbulence has developed. At small 
scales for all runs, $S_2 \sim \ell_\perp^2$, consistent 
with a smooth field in the dissipative range. At large scales, $S_2$ is 
larger for run B3 than for runs B1 and B2, a signature of the inverse 
cascade of energy and of the development of large scale structures in 
the flow. The scaling of runs B1 and B2 at intermediate scales is 
compatible with the Kolmogorov spectrum, while the scaling in run B3 is 
consistent with the $\sim k_\perp^{-2}$ energy spectrum. Note that such 
a scaling can be understood as a slow-down in the energy transfer rate 
because of interactions between waves and eddies (see e.g., 
\cite{Zhou95,Muller07,Mininni07}); such a slow-down is consistent with 
the results of the transfer function presented in the previous section. 
Considering the energy flux in the inertial range $\epsilon$ is slowed 
down by (see e.g., \cite{Iroshnikov63,Kraichnan65})
\begin{equation}
\epsilon \sim \delta u_{\ell_\perp}^2 \tau_\Omega / \tau_{\ell_\perp}^2 ,
\end{equation}
where $\tau_\Omega \sim 1/\Omega$, and 
$\tau_{\ell_\perp} \sim \ell_\perp/\delta u_{\ell_\perp}$ is the turnover 
time of eddies in the plane perpendicular to $\mbox{\boldmath $\Omega$}$;
the scaling
\begin{equation}
\delta u_{\ell_\perp}^2 \sim \ell_\perp
\end{equation}
follows.

\begin{table}
\caption{\label{table:micro}Characteristic scales and dimensionless 
         numbers of the runs in set B. $t$ is the time, $L_\parallel$ 
         and $L_\perp$ are the integral scales using respectively the 
         $E(k_\parallel)$ and $E(k_\perp)$ spectra, $\lambda$ is the 
         isotropic Taylor scale, $Ro_\lambda$ is the micro-Rossby number, 
         and $\mu = 2 \zeta_3 - \zeta_6$.}
\begin{ruledtabular}
\begin{tabular}{ccccccc}
Run & $t$ & $L_\parallel$ & $L_\perp$ & $\lambda$ & $Ro_\lambda$ & $\mu$   \\
\hline
B1 & $16$ & $1.5$         & $0.9$     & $0.29$    & $3.70$ & $0.23\pm0.01$ \\
B2 & $24$ & $0.9$         & $1.6$     & $0.31$    & $0.91$ & $0.24\pm0.01$ \\
B3 & $20$ & $2.6$         & $1.2$     & $0.50$    & $0.12$ & $0.19\pm0.02$ \\
B3 & $25$ & $2.4$         & $1.5$     & $0.55$    & $0.11$ & $0.26\pm0.02$ \\
B3 & $30$ & $2.1$         & $1.7$     & $0.59$    & $0.12$ & $0.26\pm0.05$ \\
B3 & $40$ & $1.9$         & $2.8$     & $0.53$    & $0.33$ & $0.24\pm0.02$ \\
\end{tabular}
\end{ruledtabular}
\end{table}

Figure \ref{fig:exponent} shows the scaling exponents $\zeta_p$ up to
order 6 computed in runs B1, B2, and B3. The scaling exponents are 
defined as the exponents in
\begin{equation}
S_p(\ell_\perp) \sim \ell_\perp^{\zeta_p}
\end{equation}
in the inertial range associated to the direct cascade of energy (i.e., 
for $\ell_\perp < L_F$). 
 Runs B1 and B2 behave as non-rotating turbulence, 
with Kolmogorov scaling ($\zeta_p \approx 2/3$) and intermittency
corrections (the prediction $\zeta_p = p/3$ of Kolmogorov, and the 
intermittency model of intermittency in homogeneous and isotropic 
turbulence of She and L\'ev\^eque \cite{She94} are shown in Fig. 
\ref{fig:exponent} as a reference). However, run B3 has a distinct 
behavior with $\zeta_2 \approx 1$. As time evolves in this run, and 
the energy piles up at $k_\perp \approx 1$, the second order scaling 
exponent slowly converges to this value. Low order moments follow the 
curve $\zeta_p = p/2$, but high order moments deviate from the straight 
line. The level of intermittency in the flow in all these runs can be 
measured in terms of $\mu = 2 \zeta_3 - \zeta_6$. This quantity, 
together with the integral scales of the flow (based on the parallel 
and perpendicular energy spectra), the Taylor scale, and the 
micro-Rossby number (based on the Taylor scale of the flow),
\begin{equation}
Ro_\lambda = \frac{U}{2 \Omega \lambda},
\end{equation}
are given in Table \ref{table:micro} for the runs in set B at different 
times.

\begin{figure}
\includegraphics[width=8cm]{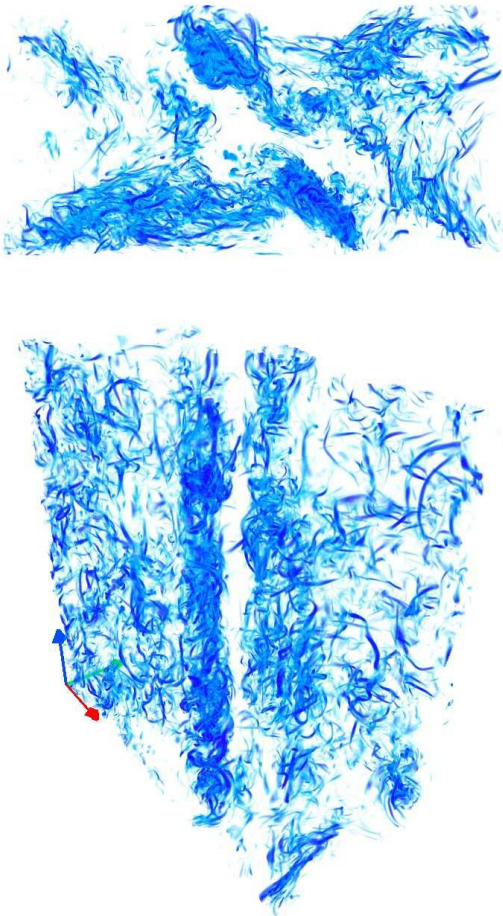}
\caption{(Color online) Three dimensional rendering of the vorticity 
intensity in a subvolume of $256 \times 512 \times 512$ grid points of 
run B2. The top view shows the subvolume in the direction of the axis 
of rotation; in the bottom view the red and blue arrows indicate 
respectively the $x$ and $z$ axis.}
\label{fig:w3D} \end{figure}

It can be seen that at late times run B3 evolves towards an anisotropic 
state in the large scales, with $L_\perp/L_\parallel \approx 1.5$. However, 
at small scales the flow seems more isotropic and at late times 
($t \approx 40$) in this run $\lambda_\perp/\lambda_\parallel \approx 0.8$. 
The micro-Rossby number in runs B1, B2, and B3 take different values 
in the range $0.11$--$3.7$. However, the value of $\mu$ is, within error 
bars, approximately the same for all the runs. As a result, the 
intermittency in the direct cascade of energy seems to be independent 
of the Rossby number $Ro$ and the micro-Rossby number $Ro_\lambda$.

Finally, Figure \ref{fig:histo01} shows the time evolution of the probability 
density function (pdf) of the longitudinal velocity increments in run B3. 
Increments in the direction perpendicular to $\mbox{\boldmath $\Omega$}$ 
were computed, and the increment was taken equal to three times the 
Kolmogorov dissipation scale $\eta$ in each run. The velocity increments 
in each run were normalized by their corresponding root mean square 
deviation $\sigma$. In agreement with the level of intermitency observed 
in the scaling exponents, the pdfs show exponential tails indicating a 
larger than Gaussian probability of large gradients to occur in the 
small scales. The amplitude of the tails of the pdfs as a function of 
$\delta u/\sigma$ does not change significantly with time. Moreover, the 
root mean square deviation $\sigma$ of the velocity increments $\delta u$ 
increases with time. So if the pdfs are plotted versus $\delta u$ (instead 
of versus $\delta u/\sigma$), the pdfs actually become wider at later times. 
This effect can be understood considering that once the inverse cascade 
of energy sets in, the total energy in the flow as a function of time 
increases.

\section{Structures}

\begin{figure}
\includegraphics[width=8cm]{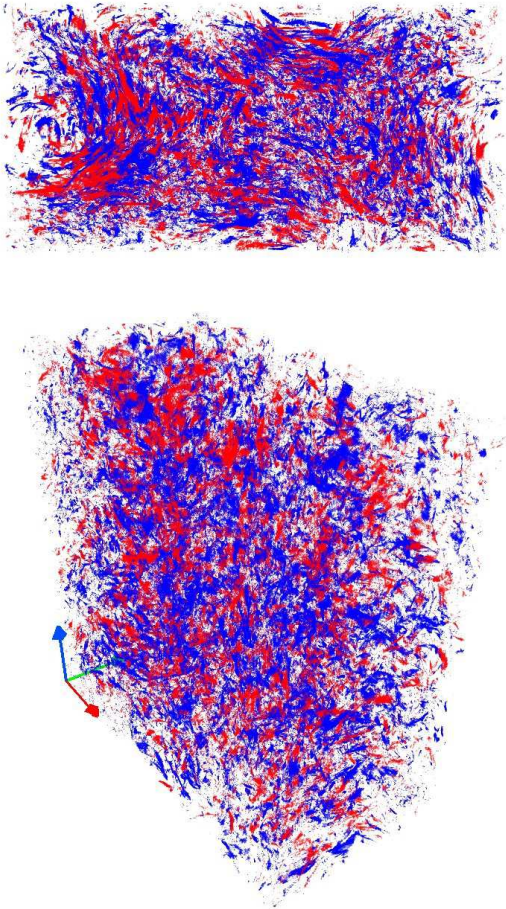}
\caption{(Color online) Three dimensional rendering of the relative 
helicity intensity in a subvolume of $256 \times 512 \times 512$ grid 
points of run B2. Blue corresponds to positive helicity, and red to 
regions with negative helicity. Only regions with 
$|\mbox{\boldmath $\omega$}\cdot{\bf u}|/(|\mbox{\boldmath $\omega$}||{\bf u}|) > 0.95$ are shown, indicating strong quenching of nonlinearities. Viewing 
points are identical to Fig. \ref{fig:w3D}.}
\label{fig:h3D} \end{figure}

The intermittency reported in the previous section in the scaling 
exponents and the pdfs of velocity increments indicates that even 
after the inverse cascade sets in, the flow develops strong velocity 
gradients in the small scales. In this section, we present 
visualizations of the flow and consider the structures that emerge.

Figure \ref{fig:w3D} shows a three dimensional rendering of the vorticity 
intensity in half of the computational domain ($256 \times 512 \times 512$ 
grid points) at late times. The top view corresponds to the subvolume in the 
direction of the axis of rotation. Only regions with strong vorticity are 
shown. Note that the flow is anisotropic and quasi-2D, as it is clear from the 
top view. In the bottom view, the development in the flow of large scale 
column-like structures can be seen. However, the columns display small 
scale structures with thin vortex filaments. These filaments seem to be 
ordered according to the large scale pattern. The presence of regions with 
strong vorticity even when the Rossby number is small enough for the 
inverse cascade of energy to develop can be expected from the 
results shown in Figs. \ref{fig:exponent} and \ref{fig:histo01}, 
linked to the intermittency of the flow.

The local relative helicity 
$\mbox{\boldmath $\omega$}\cdot{\bf u}/(|\mbox{\boldmath $\omega$}||{\bf u}|)$
in the same subvolume is shown in Fig. \ref{fig:h3D}. Unlike in isotropic 
and homogeneous turbulence, regions of strong vorticity are not correlated 
with regions of strong relative helicity. The net helicity over the entire 
box averages to zero, and locally regions of positive and negative helicity 
seem to be more isotropic and homogeneous than the other quantities 
studied. Note that the local relative helicity can be quite ubiquitously 
strong, indicative of a local quenching of nonlinear interactions.

\section{Conclusions}

In this work, we presented results of the study of the turbulent scaling 
laws and energy transfer in direct numerical simulations of rotating flows 
in periodic domains. Spatial resolutions of $256^3$ (set A) and of $512^3$ 
grid points (set B) were used, while moderate Rossby numbers (down to 
$Ro \approx 0.1$) and large Reynolds numbers (up to $Re \approx 1100$) 
were considered, with enough scale separation to observe both a direct 
and an inverse cascade of energy when the rotation was strong enough. 
Runs in set A were started from a fluid at rest, while 
runs in set B were restarted from a previous state of homogeneous 
turbulence. In the former case, for $Ro \approx 0.1$, a long transient 
was found in which the energy dissipation is small, as well as the 
energy flux to smaller scales. During this transient, the energy 
spectrum has a wide but steep spectrum, and its slope monotonously 
increases as a function of time. After turbulence sets in and the 
inverse cascade of energy develops, the energy spectrum evolves towards 
a $E \sim k_\perp^{-2}$ scaling at scales smaller than the forcing scale. 
This late time evolution is observed in both sets of runs.

At late times, the energy flux in runs A5, A6, and B3 
indicates an inverse cascade of energy in $k_\perp$ at scales larger 
than the forcing scale, and a direct cascade of energy at smaller scales. 
The net flux to small scales decreases as the Rossby number decreases, while 
the amplitude of the flux to large scales increases. No inverse cascade 
is observed in $k_\parallel$. These cascades were confirmed by the study 
of the shell-to-shell energy transfer. The direct transfer of energy 
at scales smaller than the forcing is local, although in the runs with 
small Rossby number the transfer in $k_\perp$ is significantly slowed down. 
In this direction, the energy is transfered between shells $K_\perp$ and 
$Q_\perp$ with small steps given by $|Q_\perp-K_\perp| \approx 1$. As a 
result, the direct transfer of energy in $k_\perp$ at small scales is 
mediated by interactions with the largest scale in the system, the energy 
containing eddies with $k_\perp \approx 1$. The timescale associated 
to the direct cascade in $k_\perp$ then increases, and its flux reduces.
In $k_\parallel$ the transfer is direct at all scales, and a larger 
component than in the case of non-rotating turbulence is due to 
interactions with the forcing scale. These results are in good agreement 
with phenomenological derivations of the energy spectrum in rotating 
turbulence that consider a slow down in the energy transfer rate because 
of interactions between waves and eddies \cite{Zhou95,Muller07}. The 
non-local interactions also lead to the development of anisotropies in 
the flow \cite{Waleffe93}.

The inverse cascade of energy that develops at scales larger than the 
forcing scale in runs A5, A6, and B3 is non-local, in the sense that 
the transfer of energy associated to this cascade takes place between 
disparate shells in Fourier space. At late times, the inverse transfer 
superposes with a (smaller in amplitude) direct local transfer of energy. 
This small direct transfer of energy at large scales is the result of a 
reflection at $k_\perp=1$, when the peak of energy reaches the largest 
scale in the box. Consequently, the late time evolution of simulations 
of rotating turbulence may depend on the boundary conditions used, a 
property already observed in simulations of two dimensional turbulence 
\cite{McWilliams84,Brachet86,Babiano87,Benzi87,Maltrud91}.

The study of structure functions in the direct cascade range shows that 
the second order scaling exponent for increments perpendicular to the 
rotation in runs with small $Ro$ is $\zeta_2 \approx 1$, in agreement 
with the energy spectrum. Low order moments follow the curve 
$\zeta_p = p/2$ but high order moments deviate from this law, an indication 
of intermittency. The level of intermittency in the direct cascade of 
energy, as measured by the exponent $\mu$, is the same for runs with and 
without rotation. The spontaneous formation of strong gradients in the small 
scales is further confirmed by pdfs of the velocity increments and by 
visualization of regions of strong vorticity in the flow.

More separation of scales is needed to study the intermittency in the 
inverse cascade of energy. Because of its relation to small scale 
gradients, intermittency is believed to be associated only with the 
forward cascade of energy. The intermittency phenomenon is not observed in the 
velocity field in two dimensional turbulence for which the conservation 
of vorticity leads to an inverse energy cascade to the large scales 
\cite{Benzi95,Boffetta07}, although intermittency in the vorticity 
(which cascades directly to small scales) is observed. It is unclear 
how the dual cascade of energy (towards small and large scales) in 
rotating turbulence affects the intermittency in the inverse cascade 
range. While intermittency is associated with small scale events, in 
many cases the strong events can affect the dynamics of the large 
scales, specially in systems close to criticality; as an example, 
intermittency is a possible explanation for the occurrence of extended 
minima in solar activity \cite{Charbonneau01,Mininni02}; it is also
known to affect the transport of momentum in atmospheric surface 
layers \cite{Kulkarni99}. 

\begin{acknowledgments}
The authors would like to express their gratitude to J.R. Herring and J.J. 
Tribbia for their careful reading of the manuscript. Computer time was 
provided by NCAR. PDM is a member of the Carrera del Investigador 
Cient\'{\i}fico of CONICET. AA acknowledges support from Observatoire 
de la C\^ote d'Azur and Rotary Club's district 1730. The NSF grant 
CMG-0327888 at NCAR supported this work in part. Three-dimensional 
visualizations of the flows were done using VAPOR, a software for 
interactive visualization and analysis of terascale datasets \cite{Clyne07}.
\end{acknowledgments}


\begin{thebibliography}{10}

\bibitem{Alexakis07}
{Alexakis, A., Bigot, B., Politano, H., and Galtier, S.}
\newblock Anisotropic fluxes and nonlocal interactions in magnetohydrodynamic
  turbulence.
\newblock {\em Phys.\ Rev.\ E 76\/} (2007), 056313.

\bibitem{Alexakis05b}
{Alexakis, A., Mininni, P.~D., and Pouquet, A.}
\newblock Imprint of large-scale flows on turbulence.
\newblock {\em Phys.\ Rev.\ Lett. 95\/} (2005), 264503.

\bibitem{Alexakis05}
{Alexakis, A., Mininni, P.~D., and Pouquet, A.}
\newblock Shell-to-shell energy transfer in magnetohydrodynamics. i. steady
  state turbulence.
\newblock {\em Phys.\ Rev.\ E 72\/} (2005), 046301.

\bibitem{Alexakis06}
{Alexakis, A., Mininni, P.~D., and Pouquet, A.}
\newblock On the inverse cascade of magnetic helicity.
\newblock {\em Astrophys.\ J. 640\/} (2006), 335--343.

\bibitem{Alexakis07b}
{Alexakis, A., Mininni, P.~D., and Pouquet, A.}
\newblock Turbulent cascades, transfer, and scale interactions in
  magnetohydrodynamics.
\newblock {\em New J.\ Phys. 9\/} (2007), 298.

\bibitem{Babiano87}
{Babiano, A., Basdevant, C., Legras, B., and Sadourny, R.}
\newblock Vorticity and passive-scalar dynamics in two-dimensional turbulence.
\newblock {\em J.\ Fluid Mech. 183\/} (1987), 379--397.

\bibitem{Babin96}
{Babin, A., Mahalov, A., and Nicolaenko, B.}
\newblock Global splitting, integrability and regularity of three-dimensional
  {E}uler and {N}avier-{S}tokes equations for uniformly rotating fluids.
\newblock {\em Eur.\ J.\ Mech.\ B/Fluids 15\/} (1996), 291--300.

\bibitem{Bardina85}
{Bardina, J., Ferziger, J.~H., and Rogallo, R.~S.}
\newblock Effect of rotation on isotropic turbulence: computation and modeling.
\newblock {\em J.\ Fluid Mech. 154\/} (1985), 321--336.

\bibitem{Bartello94}
{Bartello, P., M\'etais, O., and Lesieur, M.}
\newblock Coherent structures in rotating three-dimensional turbulence.
\newblock {\em J.\ Fluid Mech. 273\/} (1994), 1--29.

\bibitem{Benzi87}
{Benzi, R., Patarnello, S., and Santangelo, P.}
\newblock On the statistical properties of two-dimensional decaying turbulence.
\newblock {\em Europhys.\ Lett. 3\/} (1987), 811--818.

\bibitem{Benzi95}
{Benzi, R., and Scardovelli, R.}
\newblock Intermittency of two-dimensional decaying turbulence.
\newblock {\em Europhys.\ Lett. 29\/} (1995), 371--376.

\bibitem{Boffetta07}
{Boffetta, G.}
\newblock Energy and enstrophy fluxes in the double cascade of two-dimensional
  turbulence.
\newblock {\em J.\ Fluid Mech. 589\/} (2007), 253--260.

\bibitem{Brachet86}
{Brachet, M.~E., Meneguzzi, M., , and Sulem, P.~L.}
\newblock Small-scale dynamics of high-{R}eynolds-number two-dimensional
  turbulence.
\newblock {\em Phys.\ Rev.\ Lett. 57\/} (1986), 683--686.

\bibitem{Cambon89}
{Cambon, C., and Jacquin, L.}
\newblock Spectral approach to non-isotropic turbulence subjected to rotation.
\newblock {\em J.\ Fluid Mech. 202\/} (1989), 295--317.

\bibitem{Cambon97}
{Cambon, C., Mansour, N.~N., and Godeferd, F.~S.}
\newblock Energy transfer in rotating turbulence.
\newblock {\em J.\ Fluid Mech. 337\/} (1997), 303--332.

\bibitem{Canuto97}
{Canuto, V.~M., and Dubovikov, M.~S.}
\newblock A dynamical model for turbulence. {V}. {T}he effect of rotation.
\newblock {\em Phys.\ Fluids 9\/} (1997), 2132--2140.

\bibitem{Charbonneau01}
{Charbonneau, P.}
\newblock Multiperiodicity, chaos, and intermittency in a reduced model of the
  solar cycle.
\newblock {\em Sol.\ Phys. 199\/} (2001), 385--404.

\bibitem{Chen05}
{Chen, Q., Chen, S., Eyink, G.~L., and Holm, D.}
\newblock Resonant interactions in rotating homogeneous three-dimensional
  turbulence.
\newblock {\em J.\ Fluid Mech. 542\/} (2005), 139--164.

\bibitem{Clyne07}
{Clyne, J., Mininni, P., Norton, A., and Rast, M.}
\newblock Interactive desktop analysis of high resolution simulations:
  application to turbulent plume dynamics and current sheet formation.
\newblock {\em New J.\ Phys. 9\/} (2007), 301.

\bibitem{Embid96}
{Embid, P.~F., and Majda, A.}
\newblock Averaging over fast gravity waves for geophysics flows with arbitrary
  potential vorticity.
\newblock {\em Commun.\ Partial Diff.\ Equat. 21\/} (1996), 619--658.

\bibitem{Frisch}
{Frisch, U.}
\newblock {\em Turbulence: the legacy of A.N. Kolmogorov}.
\newblock Cambridge Univ.\ Press, Cambridge, 1995.

\bibitem{Galtier03}
{Galtier, S.}
\newblock Weak inertial-wave turbulence theory.
\newblock {\em Phys.\ Rev.\ E 68\/} (2003), 015301.

\bibitem{Greenspan}
{Greenspan, H.~P.}
\newblock {\em The theory of rotating fluids}.
\newblock Cambridge Univ. Press, Cambridge, 1968.

\bibitem{Hossain94}
{Hossain, M.}
\newblock Reduction in the dimensionality of turbulence due to a strong
  rotation.
\newblock {\em Phys.\ Fluids 6\/} (1994), 1077--1080.

\bibitem{Iroshnikov63}
{Iroshnikov, P.~S.}
\newblock Turbulence of a conducting fluid in a strong magnetic field.
\newblock {\em Sov.\ Astron. 7\/} (1963), 566--571.

\bibitem{Kraichnan65}
{Kraichnan, R.~H.}
\newblock Inertial-range spectrum of hydromagnetic turbulence.
\newblock {\em Phys.\ Fluids 8\/} (1965), 1385--1387.

\bibitem{Kulkarni99}
{Kulkarni, J.~R., Sadani, L.~K., and Murthy, B.~S.}
\newblock Wavelet analysis of intermittent turbulent transport in the
  atmospheric surface layer over a monsoon trough region.
\newblock {\em Bound.-Layer Meteor. 90\/} (1999), 217--239.

\bibitem{Maltrud91}
{Maltrud, M.~E., and Vallis, G.~K.}
\newblock Energy spectra and coherent structures in forced two-dimensional and
  beta-plane turbulence.
\newblock {\em J.\ Fluid Mech. 228\/} (1991), 321--342.

\bibitem{McWilliams84}
{Mc{W}illiams, J.~C.}
\newblock The emergence of isolated coherent vortices in turbulent flow.
\newblock {\em J.\ Fluid Mech. 146\/} (1984), 21--43.

\bibitem{Miesch00}
{Miesch, M.~S., Elliott, J.~R., Clune, J. T. T.~L., Glatzmaier, G.~A., and
  Gilman, P.~A.}
\newblock Three-dimensional spherical simulations of solar convection. {I}.
  {D}ifferential rotation and pattern evolution achieved with laminar and
  turbulent states.
\newblock {\em Astrophys.\ J. 532\/} (1999), 593--615.

\bibitem{Mininni05}
{Mininni, P.~D., Alexakis, A., and Pouquet, A.}
\newblock Shell-to-shell energy transfer in magnetohydrodynamics. ii. kinematic
  dynamo.
\newblock {\em Phys.\ Rev.\ E 72\/} (2005), 046302.

\bibitem{Mininni06}
{Mininni, P.~D., Alexakis, A., and Pouquet, A.}
\newblock Large-scale flow effects, energy transfer, and self-similarity on
  turbulence.
\newblock {\em Phys.\ Rev.\ E 74\/} (2006), 016303.

\bibitem{Mininni02}
{Mininni, P.~D., G\'omez, D.~O., and Mindlin, G.~B.}
\newblock Biorthogonal decomposition techniques unveil the nature of the
  irregularities bbserved in the solar cycle.
\newblock {\em Phys.\ Rev.\ Lett. 89\/} (2002), 061101.

\bibitem{Mininni07}
{Mininni, P.~D., and Pouquet, A.}
\newblock Energy spectra stemming from interactions of {A}lfv\'en waves and
  turbulent eddies.
\newblock {\em Phys.\ Rev.\ Lett. 99\/} (2007), 254502.

\bibitem{Morf80}
{Morf, R.~H., Orszag, S.~A., and Frisch, U.}
\newblock Spontaneous singularity in three-dimensional, inviscid,
  incompressible flow.
\newblock {\em Phys.\ Rev.\ Lett. 44\/} (1980), 572--575.

\bibitem{Muller07}
{M\"uller, W.-C., and Thiele, M.}
\newblock Scaling and energy transfer in rotating turbulence.
\newblock {\em Europhys.\ Lett. 77\/} (2007), 34003.

\bibitem{Pedlosky}
{Pedlosky, J.}
\newblock {\em Geophysical fluid dynamics}.
\newblock Springer, Berlin, 1986.

\bibitem{She94}
{She, Z.~S., and L\'ev\^eque, E.}
\newblock Universal scaling laws in fully developed turbulence.
\newblock {\em Phys.\ Rev.\ Lett. 72}, 3 (Jan 1994), 336--339.

\bibitem{Smith96}
{Smith, L.~M., Chasnov, J.~R., and Waleffe, F.}
\newblock Crossover from two- to three-dimensional turbulence.
\newblock {\em Phys.\ Rev.\ Lett. 77\/} (1996), 2467--2470.

\bibitem{Smith05}
{Smith, L.~M., and Lee, Y.}
\newblock On near resonances and symmetry breaking in forced rotating flows at
  moderate rossby number.
\newblock {\em J.\ Fluid Mech. 535\/} (2005), 111--142.

\bibitem{Smith99}
{Smith, L.~M., and Waleffe, F.}
\newblock Transfer of energy to two-dimensional large scales in forced,
  rotating three-dimensional turbulence.
\newblock {\em Phys.\ Fluids 11\/} (1999), 1608--1622.

\bibitem{Taylor37}
{Taylor, G.~I., and Green, A.~E.}
\newblock Mechanism of the production of small eddies from large ones.
\newblock {\em Proc.\ Roy.\ Soc.\ Lond.\ Ser.\ A 158}, 895 (Feb 1937),
  499--512.

\bibitem{Waleffe93}
{Waleffe, F.}
\newblock Inertial transfers in the helical decomposition.
\newblock {\em Phys.\ Fluids A 5\/} (1993), 677--685.

\bibitem{Yeung98}
{Yeung, P.~K., and Zhou, Y.}
\newblock Numerical study of rotating turbulence with external forcing.
\newblock {\em Phys.\ Fluids 10\/} (1998), 2895--2909.

\bibitem{Zeman94}
{Zeman, O.}
\newblock A note on the spectra and decay of rotating homogeneous turbulence.
\newblock {\em Phys.\ Fluids 6\/} (1994), 3221--3223.

\bibitem{Zhou95}
{Zhou, Y.}
\newblock A phenomenological treatment of rotating turbulence.
\newblock {\em Phys.\ Fluids 7\/} (1995), 2092--2094.

\end{thebibliography}

\end{document}